\documentclass[sigconf]{acmart}
\usepackage{etoolbox}
\makeatletter
\patchcmd{\maketitle}{\@copyrightspace}{}{}{}
\makeatother
\AtBeginDocument{%
  \providecommand\BibTeX{{%
    \normalfont B\kern-0.5em{\scshape i\kern-0.25em b}\kern-0.8em\TeX}}}

\copyrightyear{2024}
\acmYear{2024}
\acmDOI{}

\acmConference[Accepted for ICMHI'24]{8th International Conference on Medical and Health Informatics}{May 17--19
  2024}{Yokohama, Japan}
\acmISBN{}

\begin{document}

\title{Federated Learning and Differential Privacy Techniques on Multi-hospital Population-scale Electrocardiogram Data
}

\author{Vikhyat Agrawal}
\affiliation{%
  \institution{IIT Bombay}
  \city{Mumbai}
  \state{Maharashtra}
  \country{India}
  \postcode{400076}
}
\email{200260058@iitb.ac.in}

\author{Sunil Vasu Kalmady}
\affiliation{%
  \institution{University of Alberta}
  \streetaddress{}
  \city{Edmonton}
  \country{Canada}}
\email{kalmady@ualberta.ca}

\author{Venkataseetharam Manoj Malipeddi}
\affiliation{%
  \institution{IIT Kharagpur}
  \city{Kharagpur}
  \country{India}
}

\author{Manisimha Varma Manthena}
\affiliation{%
 \institution{University of Alberta}
 \city{Edmonton}
 \country{Canada}}

\author{Weijie Sun}
\affiliation{%
 \institution{University of Alberta}
 \city{Edmonton}
 \country{Canada}}

\author{Saiful Islam}
\affiliation{%
 \institution{University of Alberta}
 \city{Edmonton}
 \country{Canada}}

\author{Abram Hindle}
\affiliation{%
 \institution{University of Alberta}
 \city{Edmonton}
 \country{Canada}}

\author{Padma Kaul}
\affiliation{%
 \institution{University of Alberta}
 \city{Edmonton}
 \country{Canada}}

\author{Russell Greiner}
\affiliation{%
 \institution{University of Alberta}
 \city{Edmonton}
 \country{Canada}}
\renewcommand{\shortauthors}{Agrawal, et al.}

\begin{abstract}
This research paper explores ways to apply Federated Learning (FL) and Differential Privacy (DP) techniques to population-scale Electrocardiogram (ECG) data. The study learns a multi-label ECG classification model using FL and DP based on 1,565,849 ECG tracings from 7 hospitals in Alberta, Canada. The FL approach allowed collaborative model training without sharing raw data between hospitals, while building robust ECG classification models for diagnosing various cardiac conditions. These accurate ECG classification models can facilitate the diagnoses while preserving patient confidentiality using FL and DP techniques. Our results show that the performance achieved using our implementation of the FL approach is comparable to that of the pooled approach, where the model is trained over the aggregating data from all hospitals. Furthermore, our findings suggest that hospitals with limited ECGs for training can benefit from adopting the FL model compared to single-site training. In addition, this study showcases the trade-off between model performance and data privacy by employing DP during model training. Our code is available at \url{https://github.com/vikhyatt/Hospital-FL-DP}.
\end{abstract}

\keywords{Federated Learning, Differential Privacy, Electrocardiogram Data, ECG, EKG, Multi-hospital data, Healthcare, Machine Learning}
\maketitle

\section{Introduction}
Electrocardiograms (ECGs), offer critical insights into cardiac function by capturing the heart’s electrical activity, making them an indispensable tool in healthcare. As this test is non-invasive, painless, and quick, it is used to diagnose a range of cardiac conditions, including arrhythmias, conduction disorders, and other anomalies \cite{Liu2021-qa, Serhani2020-zu}. Recent studies show that ECGs can also identify non-cardiac medical conditions, including diabetes and Alzheimer’s disease \cite{Ahn2022-hm, Sun2022-nx}. Unfortunately, the manual analysis of ECG signals is a meticulous and error-prone task, even for seasoned specialists \cite{Rafie2021-cu}. Consequently, developing accurate ECG classification models is critical for the prompt and precise detection of medical conditions, which is vital for effective treatment and management. This necessity has spurred the adoption of deep learning techniques to leverage the full potential of ECG data. By automatically learning hierarchical representations from raw ECG signals, deep neural networks can uncover intricate features that might elude traditional handcrafted algorithms. However, building such models requires access to substantial and diverse datasets that represent different patient populations and medical contexts to achieve generalizable models across various clinical and demographic populations. Moreover, the acquisition of ECG data from many hospitals introduces challenges of privacy considerations. The autonomous nature of hospitals leads to many different data formats, recording systems, and ECG acquisition protocols. Additionally, certain healthcare facilities may cater exclusively to specialized clinical cohorts or serve geographically confined populations. A covariate shift between two hospitals’ data denotes a dissimilarity in patient characteristics, such as demographic features or disease prevalence, between the training dataset utilized for model development and the dataset encountered during deployment, potentially resulting in poor model performance. Concurrently, addressing privacy is important, as hospitals are bound by a mandate to safeguard patient confidentiality and uphold ethical principles \cite{aha, AMA, hipaa}. Recent research has additionally demonstrated that ECGs can be leveraged for biometric authentication, showcasing their capacity to discern and verify individuals \cite{Bras2018-ln, Melzi2023-cb}. Furthermore, given the heightened sensitivity and confidentiality of medical data, certain individuals may be reluctant to provide their data to a central data collection \cite{Ji2014-xq, Van_Zoonen2016-vo}. Thus, the establishment of collaborative, privacy-centric data-sharing frameworks assume paramount significance, with innovative methodologies such as Federated Learning (FL) and Differential Privacy (DP) emerging as promising solutions. It is essential to recognize that the FL method confers significant advantages in bolstering data security and guaranteeing privacy, as delineated by Yin et al. \cite{Yin2022-st}. This paper explores the application of FL and DP techniques on population-scale ECG data to develop accurate ECG classification models while preserving data privacy and security. 
\\
\\
Subsections \ref{1.1} and \ref{1.2} provide an overview of FL and DP, and outline the study’s objectives. Section \ref{2} then delves into related literature, exploring FL and DP applications in different healthcare contexts, especially for ECG data –to critically assess the limitations of prior work and highlight our contributions. Section \ref{3} outlines the methods we will employ, discusses the prediction task, patient characteristics, learning algorithm, and the evaluation scheme. Section \ref{4} presents the results derived from various experiments conducted related to both FL and DP. Section \ref{5} interprets our work and summarizes the entire paper.
\subsection{Federated Learning} \label{1.1}
\begin{figure}[t]
  \centering
  \includegraphics[width=\linewidth]{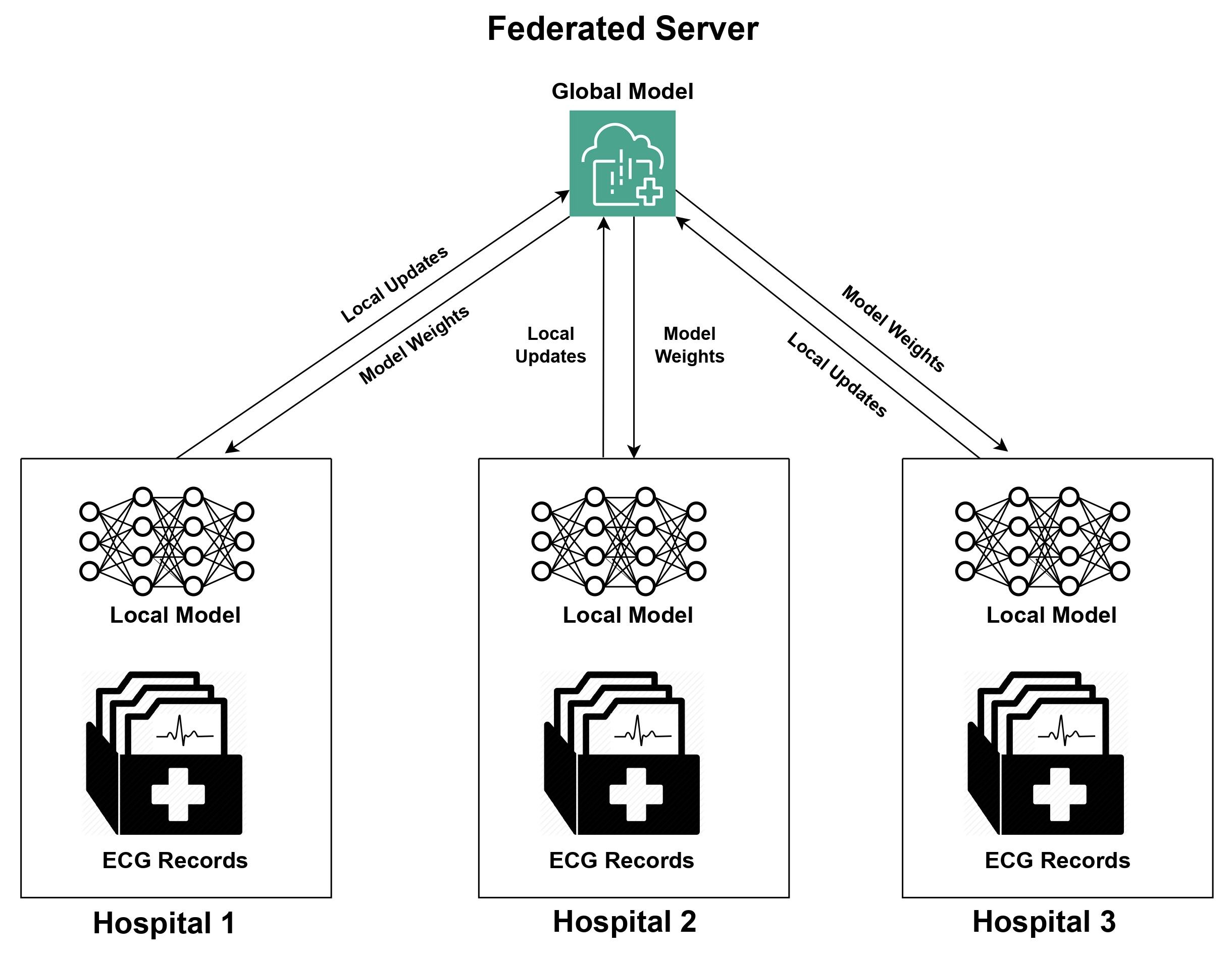}
  \caption{Diagram illustrating the Federated Learning setup for multi-hospital ECG datasets}
  \Description{}
  \label{fig:FL}
\end{figure}

Federated Learning (FL) is a pioneering paradigm designed to address the challenges associated with learning from many data sources in a way that preserves privacy \cite{pmlr-v54-mcmahan17a}. This method allows one model to be trained using data from different sources, like hospitals, without sharing private information. Each hospital helps improve that single model through minor updates while keeping sensitive details private. FL ensures that each hospital’s data stays within its boundaries, preserving data privacy and security. Figure \ref{fig:FL} depicts the FL configuration employed in our multi-hospital setup, showing FL is a multi-step process in which each hospital initially preprocesses its local ECG data, a crucial step to ensure data consistency and implement privacy-preserving techniques. Subsequently, each hospital learns its own localized ECG classification model, based on its preprocessed data. These individual models are then aggregated into a global model through techniques such as federated averaging, which appropriately weighs each hospital’s contribution \cite{pmlr-v54-mcmahan17a}. The global model is then transmitted back to the individual hospitals, where it undergoes fine-tuning in successive iterations; these “weights” are then returned to the central, until it converges on a single global ECG classification model. Through this approach, FL establishes itself as a solution for collaborative medical research across diverse healthcare institutions.
\subsection{Differential Privacy} \label{1.2}
\begin{figure}[h]
  \centering
  \includegraphics[width=\linewidth]{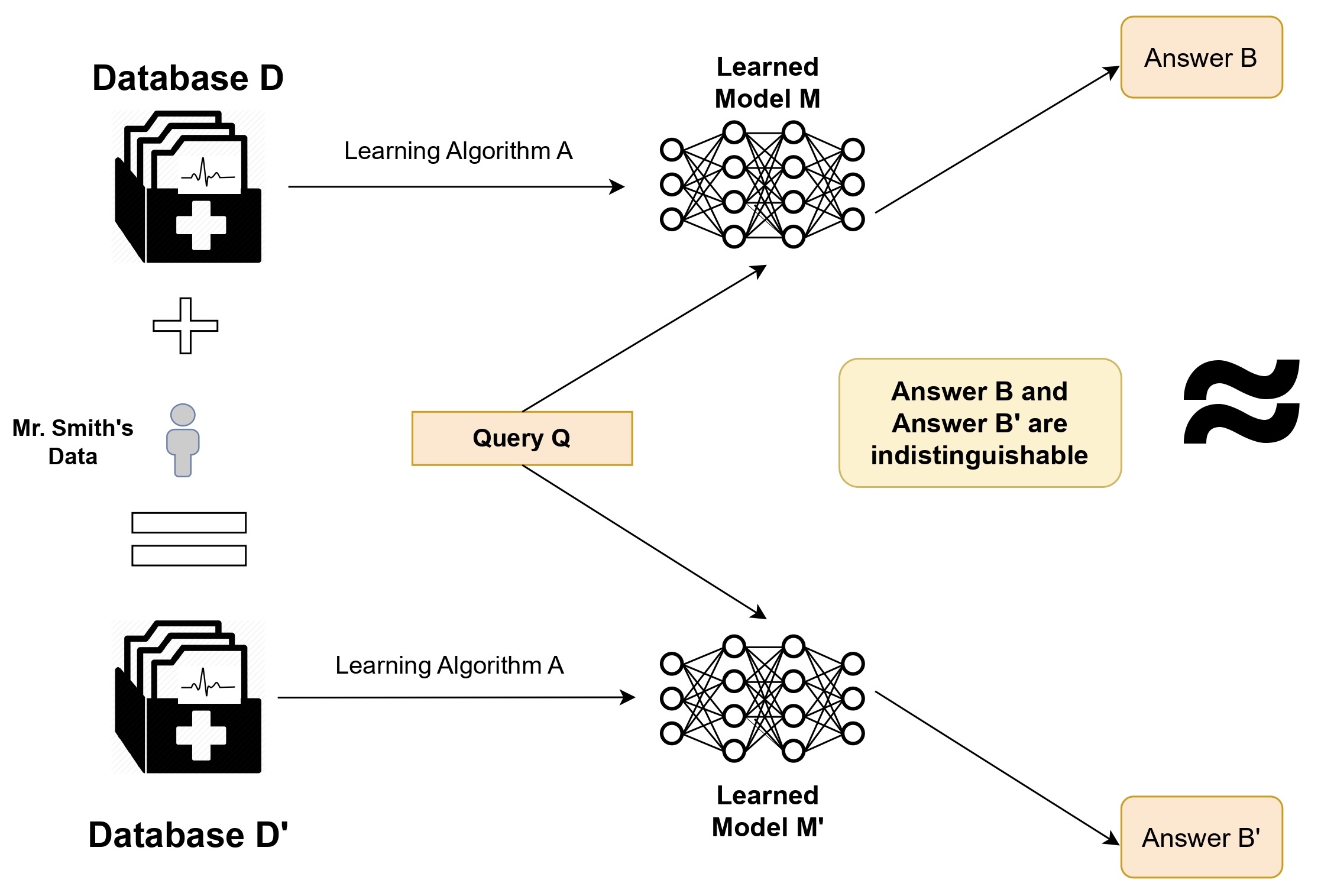}
  \caption{Diagram illustrating Differential Privacy}
  \Description{}
  \label{fig:DP}
\end{figure}
Differential privacy (DP), a fundamental concept in data privacy, has become a crucial framework for addressing the delicate balance between preserving data usefulness and safeguarding individual privacy. Initially introduced by Dwork et al. in 2006 \cite{Dwork2006-al}, this framework has since gained prominence as it offers a mathematical guarantee of privacy protection within data analysis processes. By quantifying the extent to which the inclusion or exclusion of a single individual’s data point influences the outcomes of data analyses, DP ensures that the privacy of any individual remains effectively preserved, inspiring confidence in data sharing, which facilitates collaborative medical research for advancing healthcare outcomes. Figure \ref{fig:DP} illustrates the DP setup in the context of machine learning. In this scenario, we can train 2 models using learner A: model M on database D and M’ on database D’, which is the dataset formed by adding Mr. Smith’s data to dataset D. Subsequently, we pose the same query Q to both of these trained models. If the resulting answers, B and B’, are essentially indistinguishable, then no private information about Mr. Smith can be inferred.
\subsection{Aim of Study}
Our study investigates the impact of incorporating FL and DP techniques in the analysis of ECG data obtained from 7 distinct hospitals in Alberta, Canada. Our prediction task is to develop an ECG classification model that can use a patient’s ECG to determine if that patient has zero or more of several prevalent cardiovascular and metabolic diseases. The overarching goal is to investigate whether adopting a federated approach with (versus without) differential privacy can enhance data privacy and security without significantly compromising the performance of pre-existing global models.

\section{Related Work} \label{2}
Various studies have explored the implementation of FL in healthcare applications \cite{Joshi2022-zh}. Raza et al. \cite{Raza2022-an} employed an FL approach on the MIT-BIH Dataset \cite{Moody2001-ib}, to learn a model to categorize ECGs into 5 heartbeat categories focusing on model explainability. However, it uses data from a single hospital, undermining the fundamental justification for employing FL in this scenario. In contrast, our research matches the FL objective by incorporating data aggregated from multiple hospitals. In a customized FL approach, Tang et al. \cite{Tang2021-mq} demonstrates enhanced performance in their FL model, personalized for each local node, compared to a model trained using the FedAvg algorithm. Their study would benefit from a comparative analysis across demographic groups and an inter-site performance evaluation. Our study used the same FedAvg algorithm in an inter-site analysis framework inspired by Goto et al. \cite{Goto2022-ut}, who developed FL models for hypertrophic cardiomyopathy classification by integrating ECGs and echocardiograms from 4 different hospitals, thereby setting a benchmark for assessing the utility of ECGs. We differ from the HCM-focused study \cite{Goto2022-ut} as we leverage ECG data to comprehensively diagnose various cardiovascular and metabolic diseases in patients — more details in Subsections \ref{3.1} and \ref{3.2}. Both Lin et al. \cite{9797945} and Meqdad et al. \cite{Meqdad2023-vh} propose novel FL approaches for 5-class heartbeat classification using ECGs, specifically beneficial for non-independent and identically distributed (non-IID) smart device data, with validation on the MIT-BIH dataset. The non-IID nature of the data in this context arises because the data from each node (patient or device) originates from the same patient, leading to a lack of independence and thus categorizing it as non-IID. In contrast, Baumgartner et al. \cite{Baumgartner2023-fn} learns a multi-label classification model that can detect 13 SNOMED codes using ECGs from 5 data sources \cite{moody}, emphasizing metric-focused evaluations while using FL to preserve privacy for multi-site data. Ying et al. \cite{Ying2023-gi} introduces a novel preprocessing method for ECG data by converting ECGs to images and employs FL for smart devices, incorporating a semi-supervised approach with pseudo-labeling. Nevertheless, they conducted an analysis using the MIT-BIH dataset, known for its non-IID nature and utility for patient-level data nodes. Finally, Dolo et al. \cite{10017908} proposes for the Differentially Private Stochastic Gradient Descent applied to the Federated Averaging (DPSGDFedAvg) algorithm, specifically for diabetes prediction. We adopt DPSGDFedAvg for our experiments in Section \ref{4.2}. 
\\
\\
Our study adds to the existing body of literature by introducing several advancements. An essential difference is that numerous papers utilize ECG data from the MIT-BIH dataset sourced from a single hospital. In these cases, FL is implemented on an individual patient level, treating them as nodes, and the data is non-IID. In contrast, our approach uses hospitals as nodes in our FL model, mirroring a more realistic deployment scenario. Our model utilizes data from 7 hospitals in Alberta, Canada, where each hospital contains data from numerous patients. The linkage of our ECG dataset to provincial administrative health records provides multiple diagnostic labels based on ICD-10 coding systems and our learned model can simultaneously predict the diagnosis of multiple diseases for a given ECG instance \cite{Baumgartner2023-fn, Tang2021-mq}. Our study focuses on just diseases that include at least 15,000 instances, facilitating ample training samples. We employ the FedAvg algorithm \cite{pmlr-v54-mcmahan17a} for FL, consistent with the existing literature, considering its established efficacy. Furthermore, we introduce an additional layer of data privacy by incorporating the Differential Privacy Stochastic Gradient Descent (DP-SGD) algorithm \cite{Abadi2016-fu} during FL training. The inclusion of DP-SGD differs from previous related works on similar prediction tasks and aligns with contemporary privacy norms and legal requirements. Additionally, our research conducts inter-site analysis, addressing a notable gap in some previous literature. This analysis underscores the motivation behind employing FL, particularly for hospitals with limited ECG data, thus ensuring both efficacy and data privacy simultaneously, a crucial aspect not explored in earlier studies. In addition, this study seeks to showcase the trade-off between model performance and data security by employing DP techniques in model training.

\section{Methods} \label{3}
\subsection{Data Description and Patient Characteristics} \label{3.1}
This work began with a dataset containing 2,015,808 ECG records from 260,065 patients, collected between February 2007 and April 2020, from 14 hospitals in Edmonton, Alberta, Canada. We applied the ECG data preprocessing methods outlined in the Analysis Cohort subsection of the Methods section in Sun et al. \cite{Sun2023-lp}. An ECG record was labeled with ICD-10 codes of the healthcare episode if its acquisition date fell within the timeframe of the episode. We excluded ECGs that (1) could not be linked to any episode, (2) were from patients below 18 years of age, or were of poor signal quality, leaving an analysis cohort of 1,603,109 ECGs originating from 748,773 episodes involving 244,077 patients \cite{Sun2023-lp}. In this study, we included 7 hospitals (also referred to as ‘sites’), each of which provided a minimum of 12,500 ECGs. Table \ref{tab:ecg_dist} provides the distribution of ECGs across various hospitals along with different experimental splits. The final dataset consists of 1,565,849 ECGs, from 243,128 patients across 7 hospitals. Here, each ECG is a standard 12-lead ECG tracings derived from the Philips IntelliSpace ECG system, consisting of voltage-time series, sampled at a rate of 500 Hz over a duration of 10 seconds for each of the 12 leads, resulting in a total of 500 × 10 × 12 voltage measurements per ECG. General patient characteristics have been described by Sun et al. \cite{Sun2023-lp}, and hospital-wise distributions are presented in Table \ref{tab:pat_demo}.
\begin{table}
  \caption{ECG distribution among hospitals}
  \label{tab:ecg_dist}
  \begin{tabular}{crrr}
    \toprule
    \textbf{Site} \textbackslash \textbf{Data} \textbf{Split}&\textbf{Train}&\textbf{Tuning}&\textbf{Holdout}\\
    \midrule
    Hospital 1 & 232517 & 59142 & 195383 \\
    Hospital 2 & 217578 & 54774 & 180733 \\
    Hospital 3 &  98865 & 25323 & 81938  \\
    Hospital 4 & 87201 & 21436 & 72442  \\
    Hospital 5 & 78329&20420&65035 \\
    Hospital 6 & 19769&5184&16334  \\
    Hospital 7 & 13714&3261&11291 \\
    \textbf{TOTAL} & \textbf{747973}&\textbf{194720}&\textbf{623156} \\
    
  \bottomrule
\end{tabular}
\end{table}
\\
\\
\begin{table*}
  \caption{Patient Characteristics and Prevalence of classification labels and comorbidities}
  \label{tab:pat_demo}
  \resizebox{\textwidth}{!}{%
  \begin{tabular}{crrrrrrr}
    \toprule
\textbf{Hospital}&\textbf{1}&\textbf{2}&\textbf{3}&\textbf{4}&\textbf{5}&\textbf{6}&\textbf{7}\\
    \midrule
    \hline
    \multicolumn{8}{|c|}{Patient Characteristics} \\
    \hline
    Number of ECGs&487042&453085&206126&181079&163784&41287&28266\\
    Median Age (min-max)&66 (18 - 108)&66 (18 - 108)&69 (18 - 106)&71 (18 - 108)&71 (18 - 108)&64 (18 - 105)&72 (18 - 107)\\
    Male \%&59.66&58.24&55.77&50.61&54.13&53.40&52.20\\ 
\hline
    \multicolumn{8}{|c|}{Prevalence of ICD-10 Classification Labels} \\
    \hline
   I21.1&16417 (3.37\%)&19999 (4.41\%)&4786 (2.32\%)&3131 (1.73\%)&3972 (2.43\%)&615 (1.49\%)&394 (1.39\%)\\
I21.0&14329 (2.94\%)&16840 (3.72\%)&3022 (1.47\%)&2250 (1.24\%)&2859 (1.75\%)&436 (1.06\%)&274 (0.97\%)\\
I50.0&78074 (16.03\%)&62743 (13.85\%)&23251 (11.28\%)&21970 (12.13\%)&19721 (12.04\%)&2467 (5.98\%)&3818 (13.51\%)\\
I25.10&118240 (24.28\%)&125316 (27.66\%)&33251 (16.13\%)&27869 (15.39\%)&33001 (20.15\%)&3819 (9.25\%)&3573 (12.64\%)\\
I48.9&52394 (10.76\%)&49914 (11.02\%)&14291 (6.93\%)&15671 (8.65\%)&14132 (8.63\%)&2161 (5.23\%)&2447 (8.66\%)\\
I21.4&47306 (9.71\%)&46823 (10.3\%)&25162 (12.21\%)&17054 (9.42\%)&19789 (12.08\%)&2011 (4.87\%)&2286 (8.09\%)\\
I48.0&55344 (11.36\%)&30892 (6.82\%)&17753 (8.61\%)&17041 (9.41\%)&12924 (7.89\%)&1145 (2.77\%)&1767 (6.25\%)\\
E87.5&14263 (2.93\%)&10635 (2.35\%)&3220 (1.56\%)&2788 (1.54\%)&1841 (1.12\%)&413 (1.00\%)&635 (2.25\%)\\
E11.2&22683 (4.66\%)&21745 (4.80\%)&6680 (3.24\%)&4305 (2.38\%)&4106 (2.51\%)&631 (1.53\%)&492 (1.74\%)\\
I35.0&14207 (2.92\%)&6799 (1.50\%)&1914 (0.93\%)&2066 (1.14\%)&1827 (1.12\%)&163 (0.39\%)&232 (0.82\%)\\
\hline
    \multicolumn{8}{|c|}{Prevalence of Comorbidities} \\
    \hline
    Peripheral Vascular Disease & 7184 (1.48\%) & 3393 (0.75\%) & 19582 (9.50\%) & 1504 (0.83\%) & 1262 (0.77\%) & 151 (0.37\%) & 194 (0.69\%) \\
Cerebrovascular Disease & 28453 (5.84\%) & 7684 (1.70\%) & 7992 (3.88\%) & 3516 (1.94\%) & 3525 (2.15\%) & 802 (1.94\%) & 729 (2.58\%) \\
Myocardial Infarction & 113887 (23.38\%) & 120451 (26.58\%) & 37501 (18.19\%) & 30539 (16.87\%) & 36440 (22.25\%) & 4011 (9.71\%) & 4424 (15.65\%) \\
Hypertension & 42189 (8.66\%) & 36162 (7.98\%) & 18944 (9.19\%) & 13852 (7.65\%) & 14653 (8.95\%) & 2244 (5.44\%) & 3037 (10.74\%) \\
Dementia & 7624 (1.57\%) & 8421 (1.86\%) & 3604 (1.75\%) & 5137 (2.84\%) & 3248 (1.98\%) & 149 (0.36\%) & 690 (2.44\%) \\
Chronic Pulmonary Disease & 28407 (5.83\%) & 34426 (7.60\%) & 15136 (7.34\%) & 16337 (9.02\%) & 16034 (9.79\%) & 3469 (8.40\%) & 3271 (11.57\%) \\
Renal Disease & 9799 (2.01\%) & 4466 (0.99\%) & 1781 (0.86\%) & 1718 (0.95\%) & 1634 (1.00\%) & 202 (0.49\%) & 359 (1.27\%) \\
Liver Disease & 8706 (1.79\%) & 5233 (1.15\%) & 1530 (0.74\%) & 1631 (0.90\%) & 1172 (0.72\%) & 221 (0.54\%) & 197 (0.70\%) \\
Cancer & 40575 (8.33\%) & 24215 (5.34\%) & 9437 (4.58\%) & 8294 (4.58\%) & 5887 (3.59\%) & 671 (1.63\%) & 1223 (4.33\%) \\
  \bottomrule
\end{tabular}}
\end{table*}

\subsection{Prediction Task} \label{3.2}
\begin{figure}[t]
  \centering
  \includegraphics[width=\linewidth]{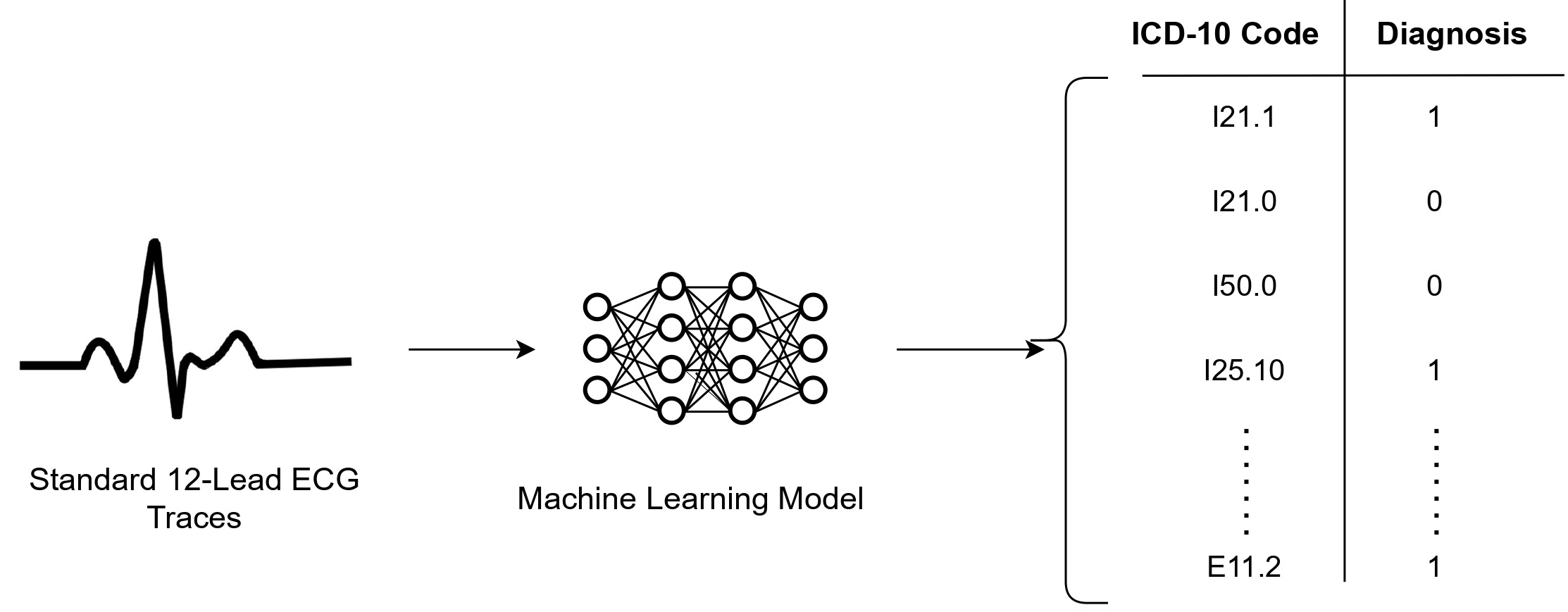}
  \caption{Diagram depicting the prediction task by representing the output using a specific example}
  \label{pred_task}
\end{figure}
\noindent
Our dataset, sourced from Alberta Health Services (AHS), contains administrative electronic health records (EHRs) and ECGs. Based on linkage of ECGs to administrative health records, we labeled each ECG with the set of ICD-10 diagnosis codes assigned to this patient. The objective of our research is a multi-label classification task that uses both ECG waveforms and demographic data to predict the probabilities of 10 specific diseases, as defined by their respective ICD-10 codes (see Appendix: Table \ref{tab:icd}). We chose these classification labels because of their significant clinical relevance to cardiovascular and metabolic diseases and their predictability using Deep Learning (DL) models, as elucidated by the research findings presented by Sun et al. \cite{Sun2023-lp}. Figure \ref{pred_task} provides a schematic of the performance system generated by our model. The model predicts each disease label independently, assigning either 0 (negative) or 1 (positive) based on the prediction probability for a given ECG tracing as input. Our dataset aggregates information from multiple hospitals, so our approach involves developing a multi-label model within a federated learning framework. We compare the pooled and federated learning models to discern the efficacy of federated learning in this context. Finally, we will compare models trained and tested independently within and across each participating hospital, evaluating their performance.

\subsection{Learning Algorithm}
For the DL model, we implemented a CNN based on the residual neural network architecture, consisting of a convolutional layer, 4 residual blocks with 2 convolutional layers per block, followed by a dense layer (total of 8,922,644 model parameters) \cite{Sun2023-lp}. We used batch normalization \cite{Ioffe2015-yu}, ReLU \cite{Agarap2018-uw}, and dropout after each convolutional layer. Each ECG instance was loaded as a 12×5120 numeric matrix. Additional features such as age and sex were passed to a 10-hidden-unit layer and concatenated with the dense layer, and finally passed to a softmax layer to produce the outputs. Binary cross-entropy was used as the loss function with the initial learning rate of $10^{-3}$, Adam optimizer \cite{Kingma2014-qn}, ReLU activation function, kernel size of 16, batch size of 512, and dropout rate of 0.2 with other hyper-parameters set to default. Models were learnt for a maximum of 50 epochs. The learning rate was reduced to $10^{-5}$ if there was no improvement in tuning loss for 7 consecutive epochs, and the learning process was stopped if loss in the tuning set did not reduce for 9 epochs. We trained all our models on the NVIDIA Driver version 418.88 with 8 Tesla V100-SXM2 GPUs and 32 GB of RAM per GPU. Each DL model took approximately 30 minutes per epoch to train.

\subsection{FL/DP Algorithm}
\subsubsection{Federated Averaging (FedAvg)} \hfill\\
We implemented the Federated Averaging (FedAvg) algorithm \cite{pmlr-v54-mcmahan17a}, a decentralized machine learning paradigm designed for training models across various hospitals while ensuring data privacy. In the FedAvg framework, each participating hospital autonomously conducts model training on its locally stored ECG data. Instead of transmitting raw data to a central server, only the model updates, in the form of gradients, are communicated. The central server aggregates these updates and computes a global model through parameter averaging. Subsequently, the resulting global model is disseminated to the participating hospitals, initiating a recurrent iterative process. FedAvg is particularly beneficial in scenarios marked by distributed ECG data across multiple sites, enabling collaborative model training without explicit data exchange.

\subsubsection{Differentially Private Stochastic Gradient Descent (DP-SGD)}\hfill\\ 
To implement differential privacy during model training, we employed the Differential Privacy Stochastic Gradient Descent (DP-SGD) algorithm \cite{Abadi2016-fu}. This optimization technique focuses on integrating differential privacy guarantees. DP-SGD extends the traditional Stochastic Gradient Descent (SGD) optimization algorithm by introducing controlled noise into gradient computations. This approach is particularly effective in mitigating the risk of extracting sensitive personal information from the ECG data. The intentional addition of noise ensures that the model’s updates maintain differential privacy, whereby the inclusion or exclusion of any single ECG has a negligible impact on the overall model parameters. DP-SGD strikes a balance between optimizing model performance and upholding the privacy of patient information within the ECG data. We have incorporated DP-SGD into our framework utilizing the Opacus library \cite{opacus}, which employs the Rényi Differential Privacy \cite{8049725} accountant as its mechanism for tracking privacy.

\subsection{Evaluation}
We randomly assigned the ECGs into the development set (60\%) and holdout set (40\%). In the development set, we used 80\% for training and 20\% for tuning the model’s performance. The proportions were maintained approximately within each hospital site (Table \ref{tab:ecg_dist}). The holdout set was kept aside to independently test the model. To mitigate the bias, we made sure that ECGs from the same patient were not used in both the development and holdout sets. For each of the 10 ICD code labels corresponding to specific medical diagnoses, we calculated the area under the receiver operating characteristic curve (AUROC) \cite{Bradley1997-fv} values based on the holdout set. Subsequently, we computed the macro AUROC score encompassing all diseases. This process was repeated 1000 times through repeated random sampling, involving 10\% of the holdout data on each occasion. Then, we calculated the mean of the 1000 macro AUROCs and determined the 95\% confidence intervals using the percentile method. We used these 95\% confidence intervals to discern statistical significance in our results.

\section{Results} \label{4}
\subsection{Characteristics of Multi-hospital Cohorts} \label{4.1}
Table \ref{tab:pat_demo} presents the distribution of age, sex, common comorbidities and diagnostic classes in the study cohort across the hospitals, highlighting their differences. Hospitals 1 and 2 consistently exhibited the highest prevalences among all participating healthcare facilities for most classification labels. The distribution of gender and age across all hospitals is notably varied. Hospitals 4, 5, and 7 showed a higher median age range of 71-72 years, which is 5 to 8 years greater compared to Hospitals 1, 2, and 4, where the median age ranged from 64 to 66 years. Similarly, Hospitals 1 and 2 had a relatively higher percentage of men (58-59\%) compared to Hospital 4, which maintained approximately equal gender distribution. All comorbidities and diagnostic classes also showed significant variations in their prevalence rates across the hospitals. Noteworthy variations are observed in the prevalence rates of specific comorbidities, with Peripheral Vascular Disease exhibiting a significantly higher prevalence in Hospital 3 (9.50\%) compared to other hospitals (with a maximum of 1.48\%). Conversely, the prevalence of Cancer, Dementia, Heart Failure (I50.0), Paroxysmal Atrial Fibrillation (I48.0), and Myocardial Infarction was conspicuously lower in Hospital 6 compared to others.
\subsection{Federated Learning}
\begin{figure}[h] 
  \centering
  \includegraphics[width=\linewidth]{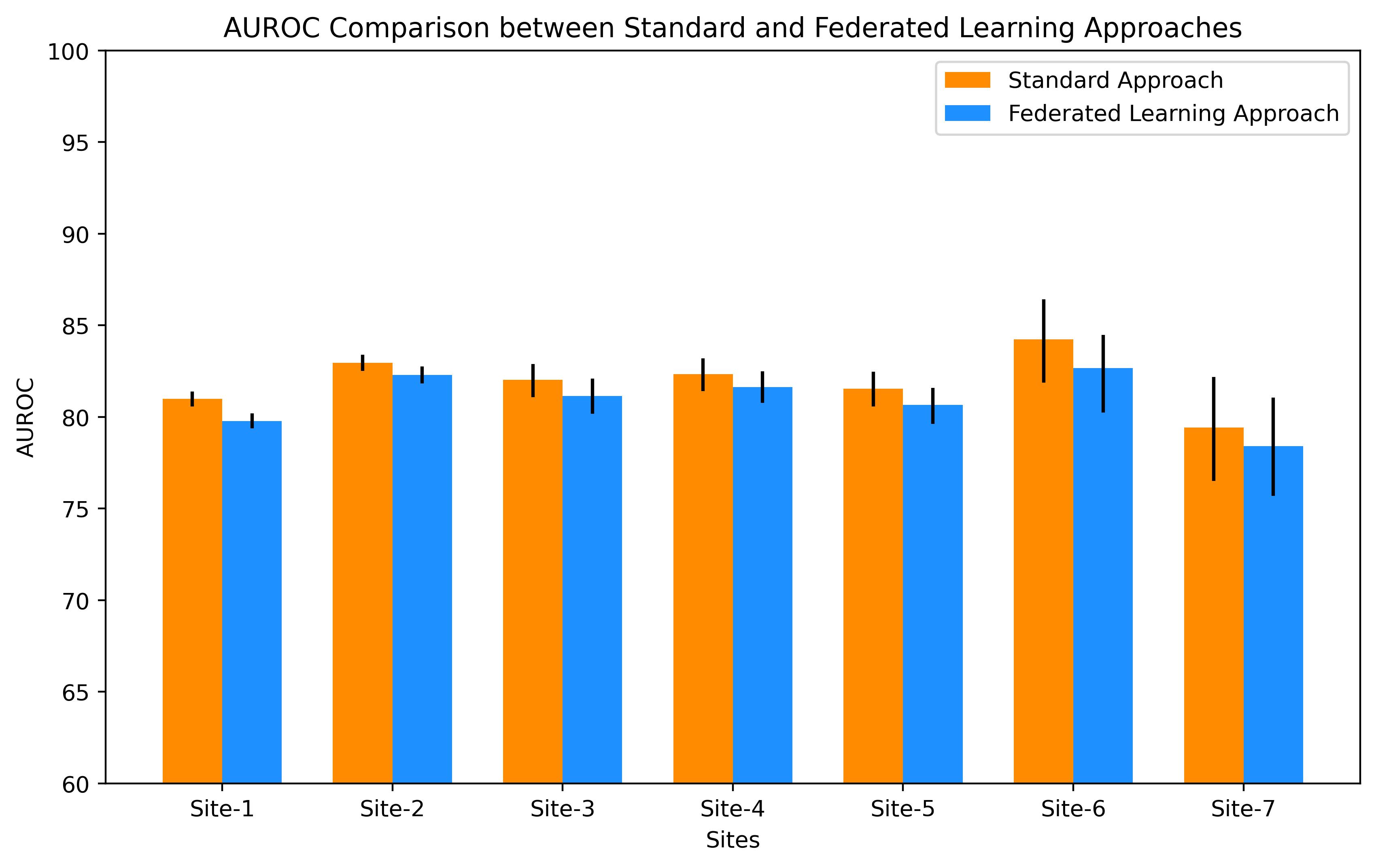}
  \caption{Comparison of Model Performance (AUROC in \%) between the Federated Learning (FL) Approach and Standard Approach for different sites. The error bars on the graph indicate the 95\% confidence intervals.
}
\label{fig:all_v_fl}
\end{figure}

Initially, we employed a conventional approach, where the entire development dataset (combining training and tuning ECGs from all the hospitals) was utilized to train a single ECG classification model. This model served as our benchmark for evaluation, as we aim to attain comparable performance using an FL setup. Subsequently, we implemented a DL model with a similar architecture within the framework of FL. We then conducted a comparative analysis of the results obtained from both models (holdout ECGs from each of the hospitals, as well as the entire holdout set). For reference, we will denote the first approach as the "pooled approach" method and the second as the "FL approach" (Table \ref{tab:standard_vs_fl}, Figure \ref{fig:all_v_fl}). We observe that specific sites, such as Site 7, exhibit low performance even with the Pooled approach. A marginal reduction in performance was observed with FL approach as opposed to the conventional data aggregation technique. However, the performance difference between the Pooled approach model and the FL approach model was statistically significant only for Site 1’s test data, while the test performance on data from all other hospitals was not statistically different. When we examined the class-wise AUROCs, we did not observe any discernible correlation between the prevalence of diseases (i.e., positive rate of labels) and their corresponding AUROC scores (Table \ref{tab:pat_demo}, Appendix: Figures \ref{standard_auroc} and \ref{fl_auroc}).
\begin{table*}
  \caption{AUROC scores of models trained in FL and Standard approach along with the respective 95\% confidence intervals}
  \label{tab:standard_vs_fl}
  \resizebox{\textwidth}{!}{%
  \begin{tabular}{ccccccccc}
    \toprule
\textbf{Model \textbackslash   Site}&\textbf{Site-1}&\textbf{Site-2}&\textbf{Site-3}&\textbf{Site-4}&\textbf{Site-5}&\textbf{Site-6}&\textbf{Site-7}& \parbox{2cm}{\textbf{Complete\\ Test Data}}\\
    \midrule
    \textbf{Standard Approach}& 80.99 [80.58,81.39]& 82.95 [82.52,83.39]& 82.02 [81.07,82.90]& 82.33 [81.42,83.20]& 81.55 [80.57,82.48]& 84.23 [81.88,86.42]& 79.43 [76.52,82.18]& 82.07 [81.58,82.50]\\
\textbf{FL approach}& 79.78 [79.37,80.20]& 82.29 [81.84,82.75]& 81.14 [80.17,82.09]& 81.63 [80.78,82.49]& 80.65 [79.63,81.58]& 82.66 [80.24,84.48]& 78.40 [75.69,81.05]& 81.03 [80.50,81.48]\\
  \bottomrule
\end{tabular}}
\end{table*}
\\
\begin{figure}[h]
  \centering
  \includegraphics[width=\linewidth]{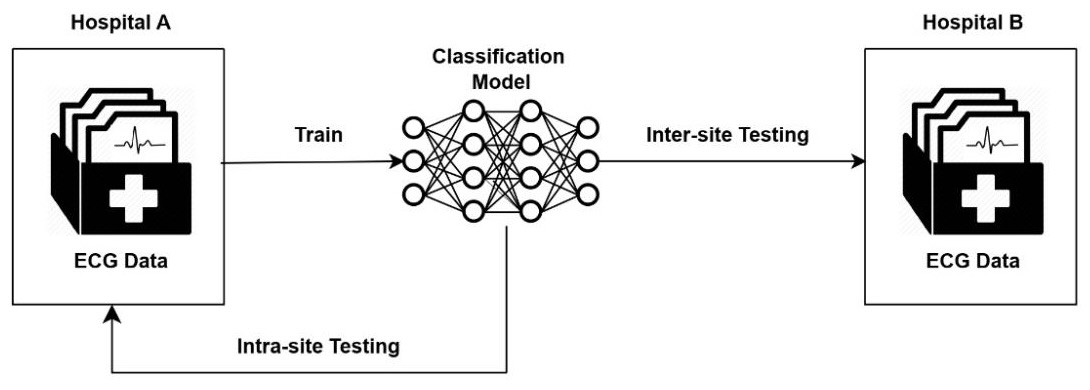}
  \caption{Diagram explaining the intra-site and inter-site testing framework. For intra-site testing, the classification model is trained and tested on Hospital A's respective train and test sets. In contrast, for inter-site testing, the model is trained on Hospital A's training data and tested on Hospital B's testing data}
    \label{fig:inter-site}
\end{figure}
\begin{table*}[h]
\centering
\caption{Table with intra-site and inter-site performance (AUROC in \%) comparisons along with corresponding 95\% confidence intervals}
\label{tab:intersite}
\resizebox{\textwidth}{!}{%
\begin{tabular}{|c|c|c|c|c|c|c|c|c|}
\hline
Train site / Test site & Site 1 (59,142) & Site 2 (54,774) & Site 3 (25,323) & Site 4 (21,436) & Site 5 (20,420) & Site 6 (5,184) & Site 7 (3,261) & Complete test data \\ 
\hline
Site 1 (232,517) & 78.87 [78.41,79.29] & 80.48 [79.97,80.98] & 79.58 [78.55,80.53] & 79.80 [78.90,80.73] & 78.99 [77.87,80.08] & 80.51 [77.87,83.03] & 77.14 [73.82,80.04] & 79.62 [79.12,80.09]\\
\hline
Site 2 (217,578) & 78.70 [78.27,79.18] & 81.61 [81.10,82.06] & 80.41 [79.40,81.33] & 80.54 [79.57,81.47] & 79.76 [78.69,80.80] & 82.39 [80.10,84.51] & 77.18 [74.02,80.13] & 80.08 [79.52,80.56] \\
\hline
Site 3 (98,865) & 77.80 [77.36,78.22] & 80.43 [79.96,80.90] & 80.28 [79.36,81.17] & 80.26 [79.34,81.19] & 79.54 [78.50,80.49] & 80.55 [77.70,83.15] & 77.36 [74.20,80.05] & 79.35 [78.86,79.82] \\
\hline
Site 4 (87,201) & 76.94 [76.45,77.39] & 79.71 [79.20,80.17] & 78.89 [77.92,79.86] & 80.04 [79.07,80.91] & 78.87 [77.79,79.99] & 81.42 [78.84,83.72] & 77.38 [74.38,80.18] & 78.57 [78.06,79.05] \\
\hline
Site 5 (78,329) &77.49 [77.05,77.91] & 80.10 [79.62,80.60] & 79.31 [78.37,80.25] & 79.87 [78.87,80.83] & 79.98 [78.95,80.97] & 80.84 [78.20,83.20] & 77.33 [74.16,80.04] & 79.12 [78.61,79.62] \\
\hline
Site 6 (19,769) & 73.04 [72.51,73.54] & 75.64 [75.05,76.20] & 75.01 [73.82,76.14] & 75.90 [78.84,83.72] & 75.19 [74.00,76.36] & 78.82 [75.51,81.53] & 73.14 [69.21,76.57] & 74.64 [74.09,75.18] \\
\hline
Site 7 (13,714) & 74.01 [73.52,74.50] & 76.70 [76.17,77.21] & 76.58 [75.46,77.60] & 76.64 [75.54,77.73] & 76.42 [75.30,77.55] & 78.66 [75.59,81.28] & 75.34 [72.43,78.44] & 75.70 [73.90,77.49] \\
\hline
Complete Train Data & 80.99 [80.58,81.39] & 82.95 [82.52,83.39] & 82.02 [81.07,82.90] & 82.33 [81.42,83.20] & 81.55 [80.57,82.48] & 84.23 [81.88,86.42] & 79.43 [76.52,82.18] & 82.07 [81.58,82.50] \\
\hline
\end{tabular}}
\end{table*} \\
In the subsequent experiment, we conducted both intra-site and inter-site testing for our ECG classification model without utilizing FL (Figure \ref{fig:inter-site}). This experiment was designed to explore how individual sites varied in their performance benefits from the FL framework, compared to local model development. Table \ref{tab:intersite} shows performance of models trained utilizing data from individual sites. \\
\\
The models trained on sites 6 and 7 demonstrated inferior performance compared to all models trained on data from the first 5 sites and tested on any of the 7 test sets. This may be attributed to the first 5 sites having significantly more number of ECG samples compared to sites 6 and 7. However, site-1 did not achieve the best performance across all test sets consistently, as the model trained on site-2 data exhibits superior performance in 4 out of the 7 test sets. Moreover, a diminishing performance trend was observed in other sites as their respective training dataset sizes decreased. Furthermore, we found that a model trained on a particular site exhibits performance that matches those trained on other sites when tested on the same site data, except for Sites 6 and 7. In other words, intra-site testing performs better than inter-site testing. For instance, a model trained on site-5 training data performs noticeably worse than other models when tested on datasets other than the site-5 testing dataset. However, when tested on site-5 data, its performance is marginally better than any of the other models. This could be attributed to covariate and distributional shifts across the sites. When comparing the performance of models trained individually for each site with those trained using the pooled approach with the complete train dataset, we found that both the pooled and FL approach models clearly surpassed their individually trained counterparts. This trend underscores the significant advantages of adopting the FL approach, as it consistently enhances the model’s performance across most sites. Particularly, hospitals with smaller sample sizes showed a marked improvement in their model performance, while hospitals with larger sample sizes exhibited a more modest impact.

\subsection{Differential Privacy}\label{4.2}
\begin{figure}[h]
  \centering
  \includegraphics[width=\linewidth]{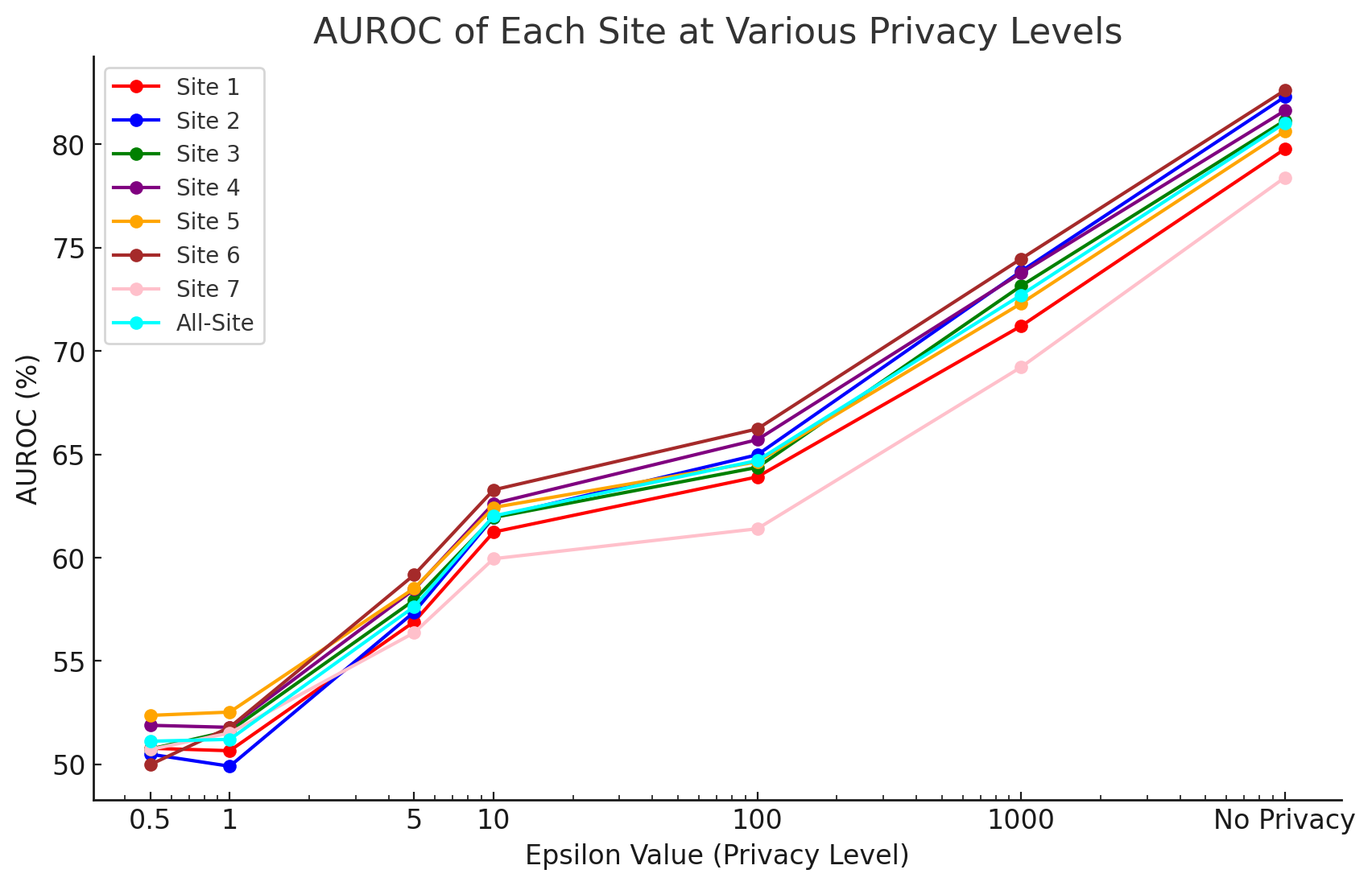}
  \caption{Plot shows model performance for different values of epsilon (tested on data from different sites)
}
\label{dp_results}
\end{figure}
\noindent
During the local training phase within FL context, we implemented DP-SGD for parameter updates at the site level. As noted in Section \ref{1.2}, the parameter $\epsilon$ in DP indicates the privacy guarantees provided by the algorithm. A smaller value of $\epsilon$ signifies a stronger privacy guarantee, implying that less information about a specific individual can be inferred from the output. As privacy is quantified using $\epsilon$, we varied this parameter to observe its influence on model performance. The model’s performance exhibited a similar diminishing trend for all of the seven sites as we reduced the $\epsilon$ value (Figure \ref{dp_results}). The model performances approached AUROC levels akin to random guessing, with values close to 0.5, for $\epsilon$ values below 1. Overall, we observed that heightened privacy protection comes at the cost of reduced model performance consistently across all sites.

\section{Discussion} \label{5}
The imperative for rigorous adherence to regulatory frameworks in healthcare, exemplified by the Health Insurance Portability and Accountability Act (HIPAA) in the United States and the General Data Protection Regulation (GDPR) in the European Union, underscores the importance of implementing FL and DP models for diagnosis tasks. Given the sensitive nature of health data and the potential for significant privacy breaches, compliance with such regulations is not merely a legal obligation but a fundamental ethical consideration. Against this backdrop, FL emerges as a pivotal research area with its inherent focus on decentralized model training and DP. This study contributes some novel findings to this end, aiming to enhance ECG ML models’ robustness and privacy assurances in compliance with prevailing legal frameworks. 
\\
\\
Our analytical framework comprises 2 distinct components: FL and DP. In the FL framework, we benchmarked our model performance against pooled models devoid of the FL technique. Our findings indicate that FL model performance registers a negligible decrease (< 2\%) across all sites compared to the pooled model trained using data from all the hospitals. During inter-site testing, we observed advantageous outcomes for hospitals with limited data under the FL paradigm. This advantage may be due to the potential inclusion of rare disease occurrences in datasets from hospitals with more extensive ECG datasets. We also note that models trained using the pooled approach on sites with many ECGs (e.g., Site 1 and Site 2) demonstrate good performance when tested across all sites, including those with fewer ECGs. Furthermore, we observe a direct correlation between the origin of training data and model performance. Models perform well when trained and tested on the same site, as displayed by a model trained on Site 5 and tested on Site 5, exhibiting comparable performance to models tested on other sites for the corresponding testing data. Moreover, our model shows increased robustness within the FL framework, owing to its training across disparate datasets with varying demographics and prevalence rates. This adaptability enhances the model’s generalization capacity, fostering resilience to variations in data.
\\
\\
Recent research has suggested alternative methodologies for conducting federated averaging on non-independent and identically distributed (non-IID) data \cite{Hong2022-yd, Tenison2022-uh, Xiao2020-sa}. We used FedAvg as a baseline model for this study to establish preliminary feasibility. We chose our prediction labels based on ICD-10 codes associated with clinically important cardiovascular and metabolic conditions, prior demonstration of predictability using ECGs \cite{Sun2022-nx}, and sufficient representation across various hospital sites. However, we acknowledge the representative nature of our label selection and highlight the need for future experiments to explore if similar trends extend to other diagnostic categories. The use of ECG machines from a single manufacturer to capture all ECG data in this study presents a potential limitation, which may restrict the diversity of instrumentation and limit the generalizability of the study’s findings. However, the examination of ECG cohorts reveals notable variations in age, sex and disease prevalence rates across the 7 hospitals despite their geographical proximity and single healthcare provider (Table \ref{tab:pat_demo}). These observed variations in clinical and demographic distributions among hospitals underscore the importance for adopting FL approaches for developing robust and equitable prediction models.
\\
\\
In the DP experiment, the DP-SGD algorithm revealed a discernible correlation between the value of $\epsilon$ and the model performance. By constraining gradients and introducing controlled noise, DP-SGD mitigates the adversarial risk on the training process. Moreover, DP-SGD can safeguard against privacy breaches, including membership inference attacks, through strategic noise injection, hindering malicious actors from inferring sensitive information. As $\epsilon$ increases, indicating a decrease in privacy, a concurrent increase in model performance is observed—an anticipated outcome. While the augmented noise levels enhance privacy safeguards, they also diminish the model’s ability to discern subtle patterns in the ECG data. Consequently, the model’s performance deteriorates when $\epsilon$ values are reduced. This observation underscores the fundamental trade-off to be considered in applications of DP: decreasing privacy constraints may augment prediction performance, whereas increasing privacy can lead to compromised model performance. We observed that the trade-off between privacy and performance becomes more pronounced with diminishing $\epsilon$ values. Therefore, striking the right balance between preserving privacy and maintaining a high level of predictability remains a complex and ongoing challenge. One potential approach for determining the optimal value of $\epsilon$ in the context of DP might be employing Membership Inference Attacks \cite{7958568} to evaluate the effectiveness of the privacy measures. Ponomareva et al. \cite{Ponomareva2023-me} posits that maintaining an epsilon value below 10 provides a substantiated privacy assurance.  Additionally, the process of computing noise to gradients can significantly increase the training time and resource requirements, which can be a practical concern in applications where real-time processing is critical. Therefore, practitioners must carefully consider these trade-offs and make informed decisions when implementing DP-SGD in their machine learning workflows. Nonetheless, the benefits of DP-SGD in terms of privacy preservation make it a compelling approach, with its limitations being areas for continued refinement in privacy-preserving machine learning.


\appendix
\section{Appendix}
\begin{figure}[h]
  \centering
  \includegraphics[width=\linewidth]{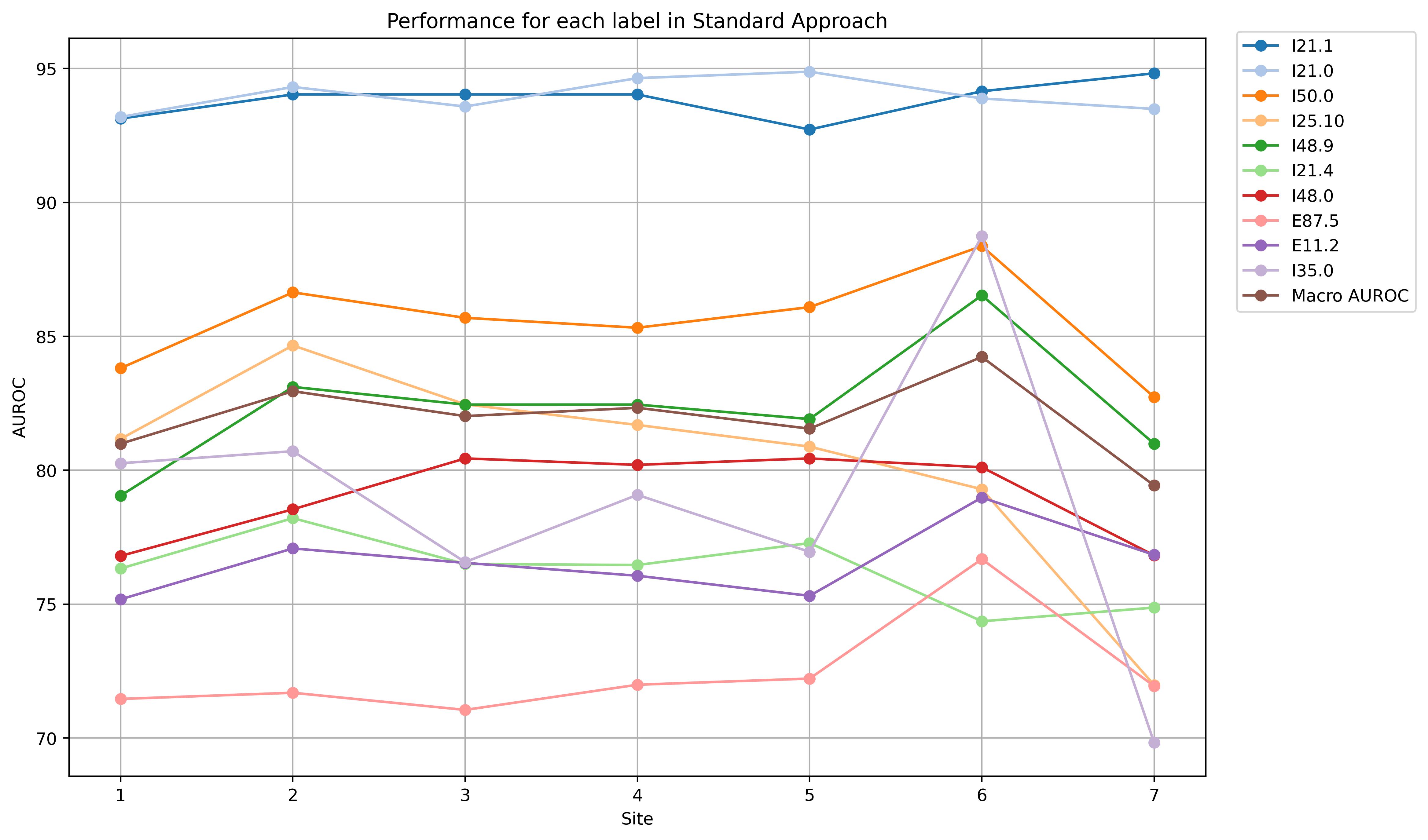}
  \caption{Plot shows model performance (AUROC in \%) for each label as well as the Macro AUROC using the Standard Approach
}
   \label{standard_auroc}
\end{figure}

\begin{figure}[h]
  \centering
  \includegraphics[width=\linewidth]{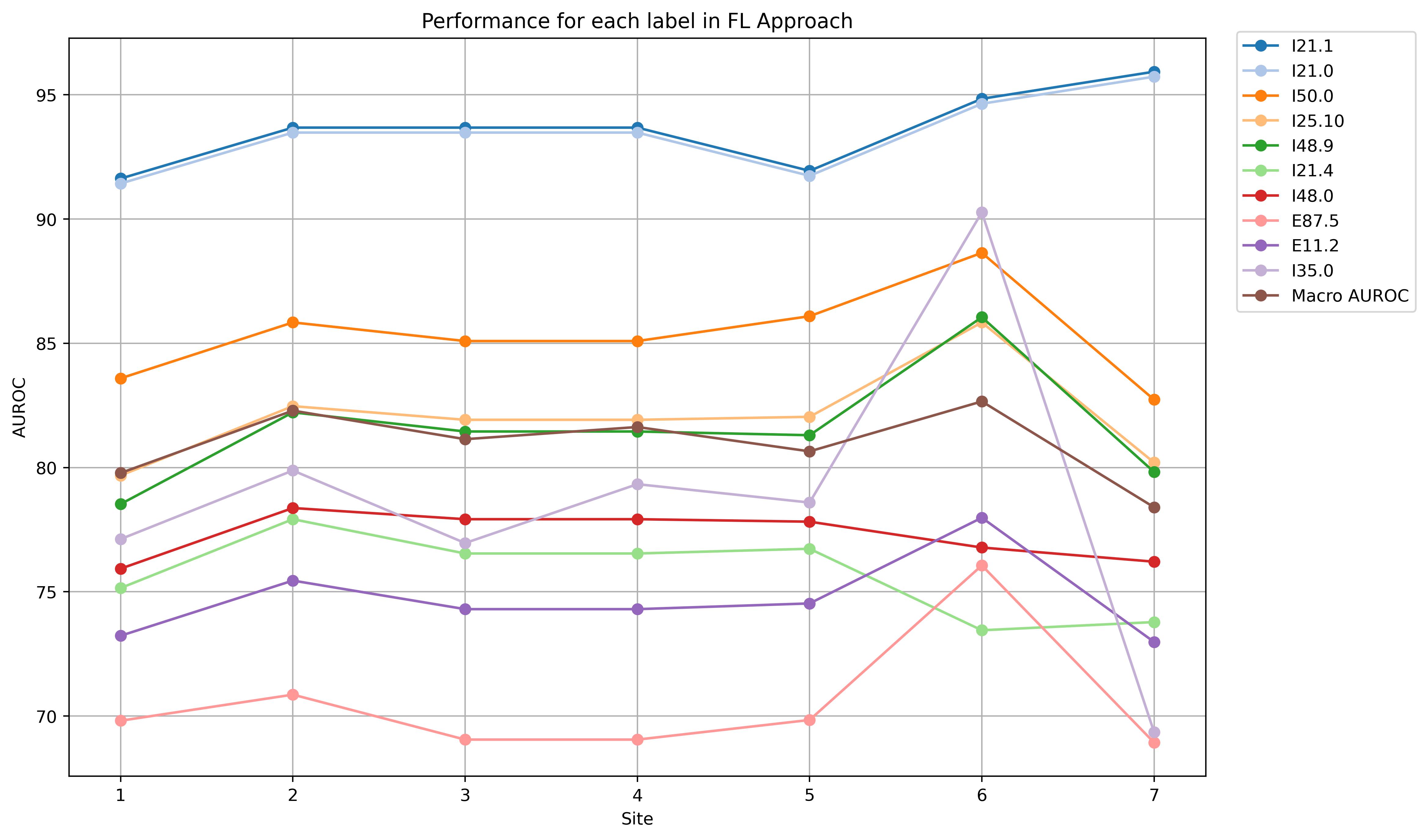}
  \caption{Plot shows model performance (AUROC in \%) for each label as well as the Macro AUROC using the Federated Learning Approach
}
\label{fl_auroc}
\end{figure}

Table \ref{tab:icd} contains the ICD-10 codes and their corresponding disease names for our classification labels. Figures \ref{standard_auroc} and \ref{fl_auroc} depict label-wise model performance across various testing sites. Figure \ref{standard_auroc} illustrates the model performance for the Standard Approach, whereas Figure \ref{fl_auroc} showcases the model performance for the Federated Learning Approach.
\begin{table}[ht]
  \caption{ICD-10 codes used and their disease names}
  \label{tab:icd}
  \begin{tabular}{cl}
    \toprule
    \textbf{ICD-10 Code} & \textbf{Disease Name}\\
    \midrule
    I21.1 &  \parbox{6.5cm}{ST elevation (STEMI) myocardial infarction \\ of inferior wall} \vspace*{7pt} \\
I21.0 & \parbox{6.5cm}{ST elevation (STEMI) myocardial infarction \\ of anterior wall} \vspace*{5pt} \\ 
I50.0 & Heart failure \vspace*{5pt} \\
I25.10 & \parbox{6.5cm}{Atherosclerotic heart disease of native coronary \\ artery without angina pectoris} \vspace*{5pt} \\ 
I48.9 & Unspecified atrial fibrillation and atrial flutter \\
I21.4 & Non-ST elevation (NSTEMI) myocardial infarction \\
I48.0 & Paroxysmal atrial fibrillation \\
E87.5 & Hyperkalemia \\
E11.2 & Type 2 diabetes mellitus with kidney complications \\
I35.0 & Nonrheumatic aortic (valve) stenosis \\
  \bottomrule
\end{tabular}
\end{table}
\\
\newpage

\end{document}